\documentclass{article}
\usepackage[onecolumn]{emulateapj}
\usepackage{psfig}
\usepackage{apjfonts}
\newcommand{\un}[2]{\mbox{\rm\thinspace #1$^{#2}$}}

\newcommand{\be}[1]{\begin{equation}\label{#1}}
\newcommand{\ee}{\end{equation}}

\newcommand{\degs}{\mbox{$^{\circ}$}}
\newcommand{\Eq}[1]{Eq.\,(\ref{#1})}

\newcommand{\Fig}[1]{Fig.\,\ref{#1}}

\newcommand{\gsim}{\mathrel{\hbox{\rlap{\lower.55ex \hbox {$\sim$}}
                   \kern-.3em \raise.4ex \hbox{$>$}}}}
\newcommand{\lsim}{\mathrel{\hbox{\rlap{\lower.55ex \hbox {$\sim$}}
                   \kern-.3em \raise.4ex \hbox{$<$}}}}

\newcommand{\Sect}[1]{Sect.\,\ref{#1}}

\newcommand{\sub}[1]{_{\rm #1}}

\newlength{\ffh}

\def\fs{\hbox{$.\!\!^{\rm s}$}}

\newcommand{\gammam}{\mbox{$\gamma\sub{m}$}}
\newcommand{\gammat}{\mbox{$\gamma\sub{t}$}}
\newcommand{\gammae}{\mbox{$\gamma\sub{e}$}}
\newcommand{\gammac}{\mbox{$\gamma\sub{c}$}}
\newcommand{\num}{\mbox{$\nu\sub{m}$}}
\newcommand{\nuc}{\mbox{$\nu\sub{c}$}}
\newcommand{\nua}{\mbox{$\nu\sub{a}$}}
\newcommand{\Cm}{\mbox{$C\sub{m}$}}
\newcommand{\Cc}{\mbox{$C\sub{c}$}}
\newcommand{\Ca}{\mbox{$C\sub{a}$}}
\newcommand{\CF}{\mbox{$C_F$}}
\newcommand{\epse}{\mbox{$\epsilon\sub{e}$}}
\newcommand{\epsB}{\mbox{$\epsilon_B$}}
\newcommand{\tday}{\mbox{$t\sub{d}$}}
\newcommand{\efiftwo}{\mbox{${\cal E}_{52}$}}
\newcommand{\me}{\mbox{$m\sub{e}$}}
\newcommand{\Mesz}{\mbox{M\'esz\'aros}}
\begin{document}

\submitted{SUBMITTED 22-APR-99 to Astrophysical Journal, in orignal form
           26-MAY-98}

\title{Physical parameters of GRB 970508 and GRB 971214 from their 
       afterglow synchrotron emission}
\author{R.A.M.J. Wijers\altaffilmark{1} and
        T.J. Galama\altaffilmark{2} 
} 
\altaffiltext{1}{Institute of Astronomy, Madingley Road, Cambridge CB3 0HA, UK
                 \& Department of Physics and Astronomy, State University
                 of New York, Stony Brook, NY 11794-3800, USA} 
\altaffiltext{2}{Astronomical Institute `Anton Pannekoek', University of 
                 Amsterdam, \& Center for High Energy Astrophysics, 
		 Kruislaan 403, 1098 SJ Amsterdam, The Netherlands}

\begin{abstract}

We have calculated synchrotron spectra of relativistic blast waves, and
find predicted characteristic frequencies that are more than an order
of magnitude different from previous calculations.  For the case of an
adiabatically expanding blast wave, which is applicable to observed
gamma-ray burst (GRB) afterglows at late times, we give expressions
to infer the physical properties of the afterglow from the measured
spectral features.

We show that enough data exist for GRB\,970508 to compute unambiguously
the ambient density, $n=0.03\un{cm}{-3}$,
and the blast wave energy per unit solid
angle, ${\cal E}=3\times10^{52}\un{erg}{}/4\pi\un{sr}{}$.  We also compute
the energy density in electrons and magnetic field. We find that they
are 12\% and 9\%, respectively, of the nucleon energy density and thus
confirm for the first time that both are close to but below equipartition.

For GRB\,971214, we discuss the break found in its spectrum by Ramaprakash
et al. (1998)\nocite{rkfkk:98}. It can be interpreted either as the
peak frequency or as the cooling frequency; both interpretations have some
problems, but on balance the break is more likely to be the cooling frequency.
Even when we assume this, our ignorance of the self-absorption frequency
and presence or absence of beaming make it impossible to constrain the
physical parameters of GRB\,971214 very well.

\end{abstract} 


   \section{Introduction}
   \label{intro}

Explosive models of gamma-ray bursts, in which relativistic ejecta
radiate away some of their kinetic energy as they are slowed down by
swept-up material, naturally lead to a gradual softening of the emission
at late times. This late-time softer radiation has been dubbed the
`afterglow' of the burst, and its strength and time dependence were
predicted theoretically (\Mesz\ and Rees, 1997)\nocite{mr:97}. Soon
after this prediction, the accurate location of GRB\,970228 by the
BeppoSAX satellite's Wide Field Cameras (Piro et al.\ 1995, Jager et
al.\ 1995)\nocite{psb:95,jhzb:95} enabled the detection of the first
X-ray and optical afterglow (Costa et al.\ 1997, Van Paradijs et
al.\ 1997)\nocite{cfhfz:97,pggks:97}. Its behavior agreed well with
the simple predictions (Wijers et al.\ 1997, Waxman 1997a, Reichart
1997)\nocite{wrm:97,waxma:97,reich:97}.

The basic model is a point explosion with an energy of order
$10^{52}\un{erg}{}$, which expands with high Lorentz factor into its
surroundings. As the mass swept up by the explosion begins to be
significant, it converts its kinetic energy to heat in a strong shock.
The hot, shocked matter acquires embedded magnetic fields and accelerated
electrons, which then produce the radiation we see via synchrotron
emission. The phenomenon is thus very much the relativistic analogue
of supernova remnant evolution, played out apparently in seconds due
to the strong time contractions resulting from the high Lorentz factors
involved.  Naturally, the Lorentz factor of the blast wave decreases as
more matter is swept up, and consequently the power output and typical
energy decrease with time after the initial few seconds of gamma-ray
emission. This produces the X-ray afterglows, which have been detected
up to 10 days after the burst (Frontera et al. 1998)\nocite{fgacf:98},
and the optical ones, which have been detected up to a year after the
burst (Fruchter et al.\ 1997, Bloom et al.\ 1998a, Castro-Tirado et al.\
1998)\nocite{flmps:97,bdkf:98,cgggp:98}.

The burst of May 8, 1997, was bright for a relatively long time and
produced emission from gamma rays to radio. This enabled a detailed
analysis of the expected spectral features of a synchrotron spectrum,
confirming in great detail that we are indeed seeing synchrotron emission,
and that the dynamical evolution of the expanding blast wave agrees with
predictions if the blast wave dynamics are adiabatic (Galama et al.\
1998a,b)\nocite{gwbgs:98,gwbgs2:98}. In principle, one can derive the
blast wave properties from the observed synchrotron spectral features. The
problem is that the characteristic synchrotron frequencies and fluxes
are taken from simple dimensional analysis in the published literature,
so they are not suitable for detailed data analysis.
 Since there are now enough data on the afterglows of a few
GRBs to derive their physical properties, we amend this situation in
\Sect{equa}, correcting the coefficients in the equations for the break
frequencies by up to a factor 10. We then use our theoretical results
to infer the physical properties of the afterglows of GRB\,970508
(\Sect{970508}) and attempt the same for
GRB\,971214 (\Sect{971214}).
We conclude with a summary of results and discuss some prospects for
future improvements in observation and analysis (\Sect{conclu}).

   \section{Radiation from an adiabatic blast wave}
   \label{equa}

      \subsection{Blast wave dynamics}
      \label{equa.dyna}

We rederive the equations for synchrotron emission from a blast
wave, to clean up some imprecisions in previous versions. Since the
dynamical evolution of the blast waves should be close to adiabatic
after the first hour or so, we specialize to the case of dynamically
adiabatic evolution. This means that the radius $r$ and Lorentz
factor $\gamma$ evolve with observer time as (Rees and \Mesz\
1992, \Mesz\ and Rees 1997, Waxman 1997a, Wijers, \Mesz, and Rees
1997)\nocite{rm:92,mr:97,waxma:97,wrm:97}
\begin{eqnarray}
    \label{eq:rt1}
         r(t) & = & r\sub{dec} (t/t\sub{dec})^{1/4} \\
    \label{eq:gamt1}
    \gamma(t) & = & \eta (t/t\sub{dec})^{-3/8}.
\end{eqnarray}
Here $\eta\equiv E/M_0c^2$ is the ratio of initial energy in the explosion
to the rest mass energy of the baryons (mass $M_0$) entrained in it, the
deceleration radius $r\sub{dec}$ is the point where the energy in the hot,
swept-up interstellar material equals that in the original explosion,
and $t\sub{dec}$ is the observer time at which the deceleration radius
is reached. Denoting the ambient particle number density as $n$, in units
of $\un{cm}{-3}$, we have
\begin{eqnarray}
    \label{eq:rdec}
         r\sub{dec} & = & \left(\frac{E}{4\pi\eta^2nm\sub{p}c^2}\right)^{1/3}  
            = 1.81\times10^{16} (E_{52}/n)^{1/3}\eta_{300}^{-2/3}\un{cm}{}\\
    \label{eq:tdec}
         t\sub{dec} & = & \frac{r\sub{dec}}{2\eta^2c} \hspace*{2.3cm}
	    = 3.35 (E_{52}/n)^{1/3}\eta_{300}^{-8/3}\un{s}{},
\end{eqnarray}
with $m\sub{p}$ the proton mass and $c$ the speed of light, and
we have normalized to typical values: $E_{52}=E/10^{52}$\,erg and
$\eta_{300}=\eta/300$. Strictly speaking, we have defined $n$ here as
$n\equiv\rho/m\sub{p}$, where $\rho$ is the ambient rest mass density.
Setting $t\sub{d}=t/1$\,day we then have, for $t>t\sub{dec}$,
\begin{eqnarray}
    \label{eq:rt2}
         r(t) & = & 2.29\times10^{17}(E_{52}/n)^{1/4} t\sub{d}^{1/4}\un{cm}{}\\
    \label{eq:gamt2}
    \gamma(t) & = & 6.65 (E_{52}/n)^{1/8} t\sub{d}^{-3/8}.
\end{eqnarray}
Note that neither $\gamma$ nor $r$ depend on $\eta$: once the blast
wave has entered its phase of self-similar deceleration, its initial
conditions have been partly forgotten. The energy $E$ denotes the initial
blast wave energy; it and the ambient density do leave their marks. It
should also be noted that these equations remain valid in an anisotropic
blast wave, where the outflow is in cones of opening angle $\theta$
around some axis of symmetry, as long as its properties are uniform
within the cone and the opening angle is greater than $1/\gamma$
(Rhoads 1998)\nocite{rhoad:98}. We should then replace $E$ by the
equivalent energy per unit solid angle ${\cal E}\equiv E/\Omega$. To
express this equivalence we shall write the normalization for this case
as ${\cal E}_{52} = {\cal E}(4\pi/10^{52}$\,erg), so we can directly
replace $E_{52}$ in all equations with ${\cal E}_{52}$ to convert from
the isotropic to the anisotropic case.

Before we can calculate the synchrotron emission from the blast wave,
we have to compute the energies in electrons and magnetic field
(or rather, summarize our ignorance in a few parameters). Firstly,
we assume that electrons are accelerated to a power-law distribution
of Lorentz factors, $N(\gamma\sub{e})\propto\gamma\sub{e}^{-p}$, with
some minimum Lorentz factor $\gamma\sub{m}$. We are ignorant of what
$p$ should be, but it can in practice be determined from the
data. The total energy in the electrons is parameterized by the ratio,
$\epsilon\sub{e}$, of energy in electrons to energy in nucleons. This
is often called the electron energy fraction, but that term is only
appropriate in the limit of small $\epsilon\sub{e}$. The post-shock 
nucleon thermal energy is $\gamma m\sub{p}c^2$, and the ratio of nucleon to
electron number densities is the same as the pre-shock value, which 
we can parameterize as $2/(1+X)$, where $X$ is the usual hydrogen  mass 
fraction. In terms of these we have
\begin{eqnarray}
    \label{eq:epsofgamm}
    \epsilon\sub{e} & \equiv & 
             \frac{n\sub{e}\langle E\sub{e}\rangle}{n\gamma m\sub{p}c^2} =
	     \frac{1+X}{2}\:\frac{m\sub{e}}{m\sub{p}}\:\frac{p-1}{p-2}\:
	     \frac{\gamma\sub{m}}{\gamma} \\
    \label{eq:gammofeps}
    \gamma\sub{m} & = & \frac{2}{1+X}\:\frac{m\sub{p}}{m\sub{e}}\:
                        \frac{p-2}{p-1}\:\epsilon\sub{e}\gamma .
\end{eqnarray}
The strength of the magnetic field in the comoving frame is parameterized by
setting the field energy density, $B^{\prime 2}/8\pi$, to
a constant fraction, $\epsilon_B$, of the post-shock nucleon energy density
$e^\prime = 4\gamma^2nm\sub{p}c^2$. (Primed quantities are measured in the
rest frame of the shocked, swept-up material; others are measured in the
frame of an observer outside the blast wave at rest relative to the
explosion center.) Consequently,
\begin{eqnarray}
   B^\prime & = & \gamma c\sqrt{32\pi nm\sub{p}\epsilon_B} \nonumber \\
   \label{eq:bprime}
            & = & 2.58\: \epsilon_B^{1/2}\efiftwo^{1/8}n^{3/8}t\sub{d}^{-3/8}
	              \: \un{G}{}.
\end{eqnarray}
From the above relations, we can express the evolution of the synchrotron
spectrum from the blast wave in terms of observable quantities and
six unknown parameters: $\efiftwo, n, X, p, \epsilon\sub{e}$, and
$\epsilon_B$. But first we need to relate the synchrotron spectrum to these
parameters.

      \subsection{Synchrotron radiation}
      \label{equa.synch}

We now derive the correct synchrotron frequencies and fluxes. These are
strictly valid only for a
uniform medium moving with a constant Lorentz factor. Real blast waves 
decelerate, of course, and have a more complicated structure behind the
shock.
The blast wave deceleration means that surfaces of
constant arrival time are no longer the ideal ellipses expected for a
constant speed of the blast wave (Rees 1966) \nocite{rees:66} and at a
given time we see contributions from gas with different Lorentz factors.
This effect has been discussed thoroughly (Waxman 1997b, Panaitescu \&
M\'esz\'aros 1998, Sari 1998)\nocite{pm:98,sari:98,waxma2:97}. The uniformity
behind the shock is also a simplification:
in reality the Lorentz factor varies from
just behind the shock to the contact discontinuity.  The density and other
parameters vary accordingly (Blandford \& McKee 1976)\nocite{bm:76}.
This effect has  not yet been treated; it is expected to be comparable
in importance to deceleration. Since both these effects are rather less
important than our corrections to the synchrotron frequencies,
we shall neglect both rather than
attempt to apply only one of them.  However, our improved treatment of
the synchrotron emission is purely local and can be incorporated into
any formalism that accounts for the varying local properties of the
shocked medium at a fixed observer time.

We assume that the electron population in any local volume has an isotropic
distribution of angles relative to the magnetic field, and that the
magnetic field is sufficiently tangled that we may average the emission
properties assuming a random mix of orientation angles between the field
and our line of sight. The radiated power per electron per unit frequency,
integrated over emission angles is
\be{eq:ppri}
   P^\prime(\nu/\nu_\perp(\gammae),\alpha) = 
   \frac{\sqrt{3}e^3B^\prime\sin\alpha}{\me c^2}\:
              F\left(\frac{\nu}{\nu_\perp\sin\alpha}\right)
	      \un{erg}{}\un{cm}{-2}\un{s}{-1}\un{Hz}{-1}\un{electron}{-1},
\ee
where $F$ is the standard synchrotron function (e.g., Rybicki \& Lightman
1979)\nocite{rl:79}, and $e$ and $m\sub{e}$ are the electron charge and mass.
$\alpha$ is the angle between the electron velocity and the magnetic field,
and 
\be{eq:nuperp}
   \nu_\perp(\gammae) = \frac{3\gammae^2eB^\prime}{4\pi\me c}.
\ee
(i.e., the traditional characteristic synchrotron frequency
equals $\nu_\perp\sin\alpha$ in our notation.)
Next, we define the 
isotropic synchrotron function $F\sub{iso}$ by averaging over an isotropic
distribution of $\alpha$. Setting $x_\perp(\gammae)=\nu/\nu_\perp(\gammae)$,
we get
\begin{eqnarray}
   \label{eq:pisopri}
   P^\prime\sub{iso}(x_\perp) & = &
        \frac{\sqrt{3}e^3B^\prime}{\me c^2}\:
	F\sub{iso}(x_\perp)\:\un{erg}{}\un{cm}{-2}\un{s}{-1}\un{Hz}{-1}\un{electron}{-1}    \\
   \label{eq:Fiso}
          F\sub{iso}(x_\perp) & = & 
	  \int_0^{\pi/2}{\rm\,d}\alpha\:\sin^2\alpha F(x_\perp/\sin\alpha)
\end{eqnarray}
(We have made use of the symmetry of $\sin\alpha$ to absorb a factor 1/2
into confining the integral to the first quadrant. The apparent singularity
at $\alpha=0$ poses no problems because $F$ decreases exponentially for
large values of the argument.)
Note that most calculations of blast wave spectra assume that the
spectrum peaks at frequency $\gammae^2eB^\prime/\me c$. Due to the neglect
of the factor $3/4\pi$ and the fact that $F(x)$ peaks at $x=0.28587$ and
$F\sub{iso}(x)$ at $x=0.22940$, this estimate leads to quite erroneous
inferences about blast wave properties.
%
%

\begin{figure}[ht]
\centerline{\psfig{figure=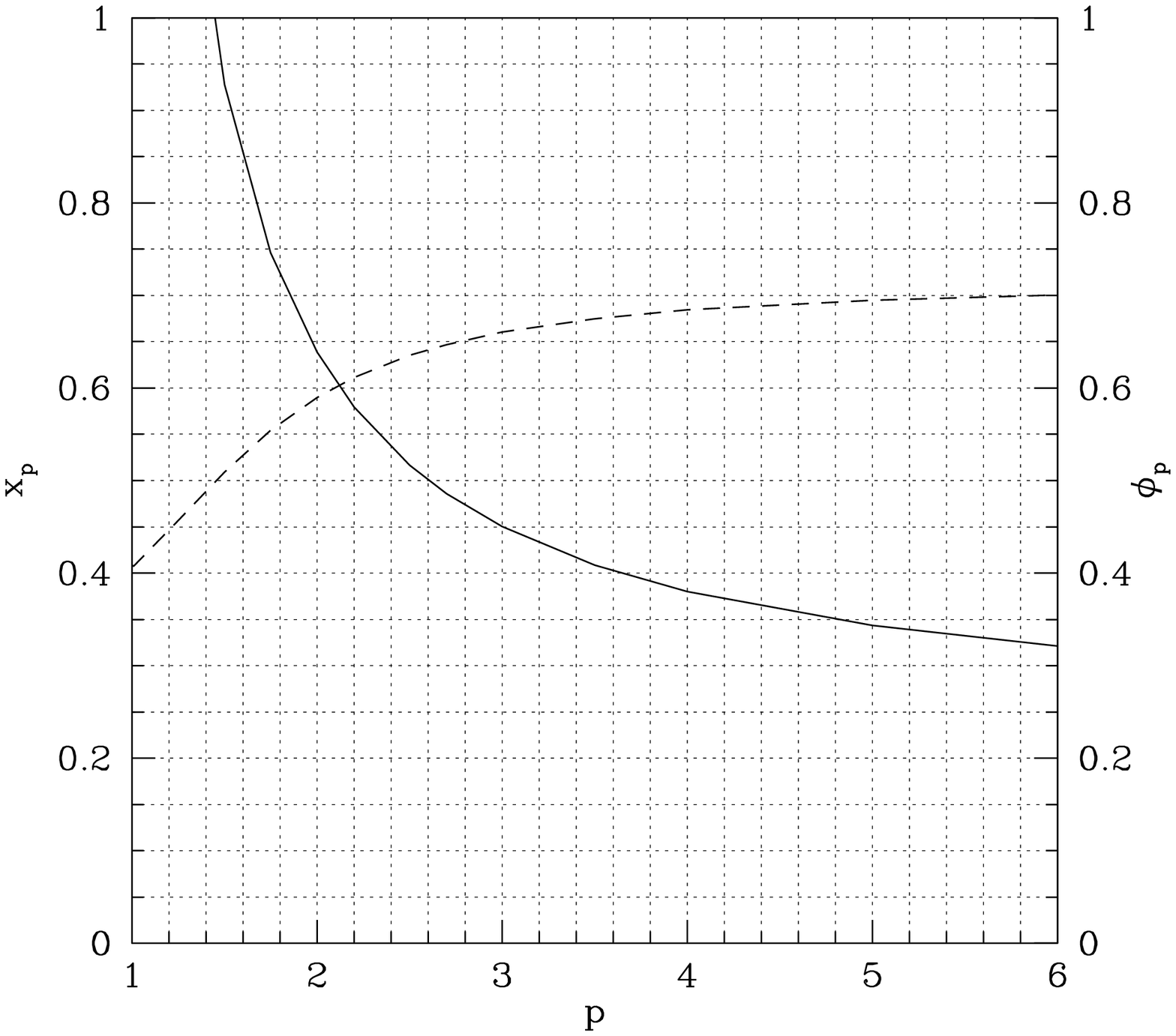,width=8.8cm}}
\caption[]{The dimensionless location $x_p$ (solid) and dimensionless
	   peak flux $\phi_p$ (dashed) of a synchrotron spectrum from
	   a power law of electrons, as a function of the power law index, $p$
	   of the electron energy distribution.
	   \label{fi:xpphip}
	   }
\end{figure}

Finally, we must average the emission over a distribution of
electron energies. We assume a simple power-law probability
distribution of electrons
between extreme values $\gammam$ and $\gammat$:
\begin{eqnarray}
   \label{eq:Pgam}
   f(\gammae) & = & \frac{f_0}{\gammam}\left(\frac{\gammae}{\gammam}\right)^{-p}
                    \hspace{1cm} \gammam\le\gammae\le\gammat \\
   f_0        & = & \frac{p-1}{1-(\gammat/\gammam)^{1-p}}
\end{eqnarray}
Now let $x=\nu/\nu_\perp(\gammam)$. Then the average
power per electron becomes
\begin{eqnarray}
   \label{eq:pPLpri}
   P^\prime\sub{PL}(x) & = &
        \frac{\sqrt{3}e^3B^\prime}{\me c^2}\:
	F\sub{PL}(x)\:\un{erg}{}\un{cm}{-2}\un{s}{-1}\un{Hz}{-1}\un{electron}{-1}    \\
   \label{eq:FPL}
          F\sub{PL}(x) & = & \frac{f_0}{2} x^{-\frac{p-1}{2}}
	  \int_{x\gammam^2/\gammat^2}^x{\rm\,d}u\:u^\frac{p-3}{2} F\sub{iso}(u),
\end{eqnarray}
in which we have transformed integration variable from $\gammae$ to
$u\equiv x\gammam^2/\gammae^2$. The last equation shows the familiar
result that for $1\lsim x\lsim \gammat^2/\gammam^2$ the spectrum from a
power law of electrons is itself a power law. Since this region is known to
extend over many decades in GRB and afterglow spectra, we quote numerical
results for the case $\gammat\gg\gammam$, for which the quoted
results are independent
of $\gammat$. The most easily identified
point in the spectrum is its dimensionless maximum, $x_p$, and the 
dimensionless flux at this point,
$F\sub{PL}(x_p)\equiv\phi_p$; their dependence on $p$ is shown
in \Fig{fi:xpphip}.  Both now depend on the electron energy 
slope $p$. This defines the
first two numbers that we can measure in the spectrum:
\begin{eqnarray}
     \nu_m^\prime & = & x_p\nu_\perp(\gammam) = \frac{3x_p}{4\pi}\:
                \frac{\gammam^2eB^\prime}{\me c} \\
     \label{eq:Pnumprim}
     P_{\nu_m}^\prime & = & \phi_p \frac{\sqrt{3}e^3B^\prime}{\me c^2}
\end{eqnarray}

The calculation of the break frequency $\nuc$ that separates radiation
from slowly and rapidly cooling electrons (Sari, Piran, and Narayan 1998)
is somewhat more difficult, because the cooling rate depends on both
$\gammae$ and pitch angle $\alpha$.
However, since the cooling and the emission are both dominated
by $\alpha=\pi/2$, we may estimate the break as the peak of $F(x)$ for
the value of $\gammae$ where the cooling time for electrons with $\alpha=\pi/2$
equals the expansion time, $t$:
\begin{eqnarray}
   \label{eq:gammac}
      \gammac & = & \frac{6\pi\me c}{\sigma\sub{T}\gamma B^{\prime 2}t} \\
      \nuc^\prime &= & 0.286 \frac{3}{4\pi}\frac{\gammac^2eB^\prime}{\me c}
\end{eqnarray}
In order to transform frequency and power from the rest frame of
the emitting material to our frame, we note that the emission is
isotropic in the rest frame by assumption. It is then trivial
to compute the angle-average Doppler factors (see Rybicki \&
Lightman 1979, Ch.4)\nocite{rl:79}. For the received power, we
find $P=\gamma^2(1+\beta^2/3)P^\prime$, which we shall simplify
to $P=4\gamma^2P^\prime/3$ in keeping with the fact that our whole
treatment is done in the ultrarelativistic limit, $\beta\rightarrow 1$.
Similarly, the intensity-weighted mean change in any frequency is $\nu =
4\gamma\nu^\prime/3$. Consequently, the appropriate mean of a power per
unit frequency will transform as $P_\nu=\gamma P_\nu^\prime$. Of course,
the spectrum also gets broadened, but that will not affect the locus of
characteristic frequencies significantly.

The synchrotron 
self-absorption frequency is usually set at the point where $\tau_\nu=0.35$.
Using the co-moving width of the shock, $\Delta r^\prime=r/4\gamma$, and
the expression for the synchrotron absorption coefficient (Rybicki \&
Lightman 1979)\nocite{rl:79}, we get
\begin{equation}
   \label{eq:nua}
   \nua = 4\gamma\nua^\prime/3 = 2.97\times10^8
	    {\scriptstyle \left(\frac{p+2}{p+\frac{2}{3}}\right)}^{3/5}
	    {\scriptstyle \frac{(p-1)^{8/5}}{p-2}} (1+X)^{8/5} n^{3/5}
            \epse^{-1} \epsB^{1/5}\efiftwo^{1/5}(1+z)^{-1}\:\un{Hz}{}
\end{equation}
where we have used equations \ref{eq:rt2}--\ref{eq:bprime}
for the blast wave dynamics to express
$\nua$ in terms of the unknowns we try to solve for, and added the correction
for redshift, i.e.\ the equation in this form relates the observed frequency
on Earth to the properties of the blast wave measured by a local observer
at rest relative to the center of the explosion.  Note that the
self-absorption frequency in this simplest form is time-independent.
We now also translate the other two frequencies into practical form:
\begin{eqnarray}
   \label{eq:num}
   \num = 4\gamma\num^\prime/3 & = & 5.73\times10^{16} x_p
          {\scriptstyle \left(\frac{p-2}{p-1}\right)}^2\epse^2\epsB^{1/2}
          \efiftwo^{1/2} (1+X)^{-2} (1+z)^{1/2}\tday^{-3/2} \:\un{Hz}{} \\
   \label{eq:nuc}
   \nuc = 4\gamma\nuc^\prime/3 & = & 1.12\times10^{12} \epsB^{-3/2}
           \efiftwo^{-1/2} n^{-1} (1+z)^{-1/2} \tday^{-1/2}\:\un{Hz}{}
\end{eqnarray}
Note the non-trivial redshift dependence of both, which stems from the fact 
that $\tday$ is also measured on Earth and therefore redshifted. The 
observed flux at $\num$ can be obtained by noting that our assumption of
uniformity of the shocked material means that all swept-up electrons since
the start contribute the same average power per unit frequency at $\num$
(at any frequency, in fact), 
which is given by \Eq{eq:Pnumprim}. Adding one factor of $\gamma$ to transform
to the lab frame and accounting for the redshift, we have:
\be{eq:Fnumobs1}
    F_{\num} = \frac{N\sub{e}\gamma P_{\num}^\prime (1+z)}{4\pi d_L^2},
\ee
where $N\sub{e}$ is the total number of swept-up electrons, related to the
blast wave parameters by $N\sub{e}=\frac{4\pi}{3} r^3 n (1+X)/2$. The
luminosity distance depends on cosmological parameters, and for an $\Omega=1,
\Lambda=0$ universe, which we shall adopt here, is given by 
$d_L=2c(1+z-\sqrt{1+z})/H_0$. Consequently, 
\be{eq:Fnumobs2}
    F_{\num} = 1.15 \frac{h_{70}^2}{(\sqrt{1+z}-1)^2}\phi_p (1+X)
             \efiftwo n^{1/2} \epsB^{1/2}\:\un{mJy}{},
\ee
where $h_{70}= H_0/70\un{km}{}\un{s}{-1}\un{Mpc}{-1}$.

Equations \ref{eq:nua}, \ref{eq:num}, \ref{eq:nuc}, and \ref{eq:Fnumobs2}
now are four independent relations between the four parameters of
interest \efiftwo, $n$, \epse, and \epsB. This means we can solve for all
parameters of interest if we have measured all three break frequencies
(not necessarily at the same time) and the peak flux of the afterglow. In
addition this requires us to know the redshift of the burst, the electron
index $p$, and the composition parameter, $X$, of the ambient medium.
Note that multiple measurements of the same break at different times serve
to test the model assumptions, but do not provide extra constraints on
the parameters, since validity of the model implies that any of the four
key equations is satisfied for all time if it is satisfied once. We therefore
define the constants
$C\sub{a}\equiv\nua/\nua_*$, $C\sub{m}\equiv\num\tday\sub{m}^{3/2}/\num_*$,
$C\sub{c}\equiv\nuc\tday\sub{c}^{1/2}/\nuc_*$, and $C_F=F_{\num}/F_{\num *}$.
Here starred symbols denote the numerical coefficients in each of the four
equations, and times denote the time at which the quantity in question 
was measured. Rearranging the four equations then yields
\begin{eqnarray}
   \label{eq:efiftwo}
   \efiftwo & = & \Ca^{-\frac{5}{6}}
              \Cm^{-\frac{5}{12}}
              \Cc^{\frac{1}{4}}
              \CF^{\frac{3}{2}}
	      x_p^\frac{5}{12}\phi_p^{-\frac{3}{2}}
	      {\scriptstyle (p-1)}^\frac{1}{2}
	      {\scriptstyle \left(\frac{p+2}{p+\frac{2}{3}}\right)}^\frac{1}{2}
	      (1+X)^{-1}(1+z)^{-\frac{1}{2}}
	      \left(\frac{\sqrt{1+z}-1}{h_{70}}\right)^3 \\
   \label{eq:epse}
   \epse    & = & \Ca^{\frac{5}{6}}
              \Cm^{\frac{11}{12}}
              \Cc^{\frac{1}{4}}
              \CF^{-\frac{1}{2}}
	      x_p^{-\frac{11}{12}}\phi_p^{\frac{1}{2}}
	      {\scriptstyle \frac{(p-1)^\frac{1}{2}}{p-2} }
	  {\scriptstyle \left(\frac{p+2}{p+\frac{2}{3}}\right)}^{-\frac{1}{2}}
	      (1+X)(1+z)^{\frac{1}{2}}
	      \left(\frac{\sqrt{1+z}-1}{h_{70}}\right)^{-1} \\
   \label{eq:epsB}
   \epsB    & = & \Ca^{-\frac{5}{2}}
              \Cm^{-\frac{5}{4}}
              \Cc^{-\frac{5}{4}}
              \CF^{\frac{1}{2}}
	      x_p^{\frac{5}{4}}\phi_p^{-\frac{1}{2}}
	      {\scriptstyle (p-1)^\frac{3}{2} } 
	    {\scriptstyle \left(\frac{p+2}{p+\frac{2}{3}}\right)}^{\frac{3}{2}}
	      (1+X)(1+z)^{-\frac{5}{2}}
	      \left(\frac{\sqrt{1+z}-1}{h_{70}}\right) \\
   \label{eq:n}
       n    & = & \Ca^{\frac{25}{6}}
              \Cm^{\frac{25}{12}}
              \Cc^{\frac{3}{4}}
              \CF^{-\frac{3}{2}}
	      x_p^{-\frac{25}{12}}\phi_p^{\frac{3}{2}}
	      {\scriptstyle (p-1)^{-\frac{5}{2}} }
	    {\scriptstyle \left(\frac{p+2}{p+\frac{2}{3}}\right)}^{-\frac{5}{2}}
	      (1+X)^{-1}(1+z)^{\frac{7}{2}}
	      \left(\frac{\sqrt{1+z}-1}{h_{70}}\right)^{-3}
\end{eqnarray}
The last factor in each of these stems from the specific 
cosmological model adopted, and has entered the solution only via
\Eq{eq:Fnumobs1}. To generalize to any cosmology, all that is needed is to
replace $(\sqrt{1+z}-1)/h_{70}$ in the above equations by 
$(d_L/8.57\un{Gpc}{})/\sqrt{1+z}$.

   \section{Observed and inferred parameters of GRB\,970508}
   \label{970508}

GRB\,970508 was a moderately bright $\gamma$-ray burst (Costa et
al. 1997, Kouveliotou et al.  1997)\nocite{cfpsc:97,kbpfm:97}. It
was detected on May 8.904 UT with the Gamma-Ray Burst Monitor
(GRBM; Frontera et al. 1991)\nocite{fetal:91}, and with the Wide
Field Cameras (WFCs; Jager et al. 1995)\nocite{jhzb:95} on board
the Italian-Dutch X-ray observatory BeppoSAX (Piro, Scarsi, \&
Butler 1995)\nocite{psb:95}.  Optical observations of the WFC error
box (Heise et al. 1997)\nocite{hzcps:97}, made on May 9 and 10,
revealed a variable object at RA = $06^{\rm h}53^{\rm m}49\fs2$,
Dec = +79\degs16\arcmin19\arcsec (J2000), which showed an increase by
$\sim$1 mag in the V band (Bond 1997)\nocite{bond:97}.  BeppoSAX Narrow
Field Instrument observations revealed an X-ray transient (Piro et
al. 1997)\nocite{pcfsa:97} whose position is consistent with that of the
optical variable, and Frail et al. (1997)\nocite{fknft:97} found
the first GRB radio afterglow for GRB\,970508; the radio source position
coincides with that of the optical source (Bond 1997)\nocite{bond:97}.

The spectrum of the optical variable showed absorption lines at redshifts
0.77 and 0.835, indicating that 0.835 is the minimum redshift of the
afterglow (Metzger et al.  1997)\nocite{mdksa:97}. Subsequently, an
[O\,II] emission line with $z=0.835$ was also found in the host's spectrum
(Bloom et~al.\ 1998a)\nocite{bdkf:98}, which is often associated with
star forming regions in galaxies. A faint underlying galaxy or star 
forming region is inferred to indeed exist from a levelling off of
the light curve after 6--11 months (Bloom et~al. 1998a, Castro-Tirado
et al.\ 1998)\nocite{bdkf:98,cgggp:98}. It must be very compact,
since the HST limits on an extended object underlying the GRB are
fainter than the magnitude inferred from the light curve (Fruchter
1998)\nocite{fruch:98}. It is therefore almost certain that the compact
nebula is the source of the [O\,II] line, and therefore also of the majority
of the absorption lines. Given its compactness, a chance location of
the burst far behind it is unlikely, and we shall assume that the burst
occurred in this nebula, i.e.\ its redshift is 0.835.

\begin{figure}[ht]
\centerline{\psfig{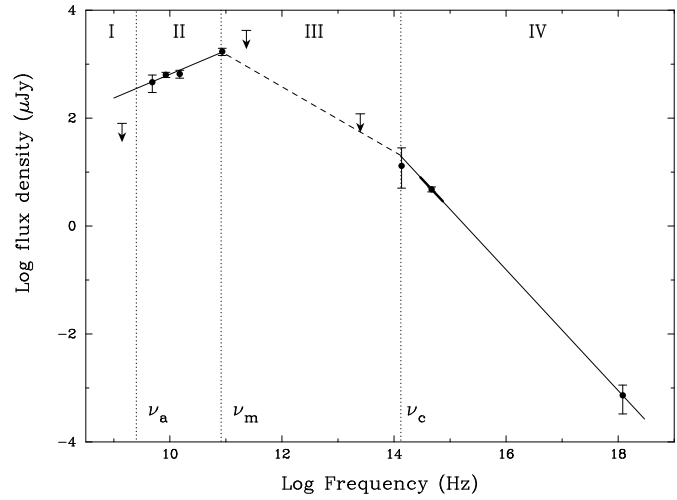}}
\caption[]{The X-ray to radio spectrum of GRB 970508 on May 21.0 UT (12.1
           days after the event) from Galama et al.\ (1998b).
           Indicated
           are the inferred values of the break frequencies $\nu_{\rm a}$,
           $\nu_{\rm m}$ and $\nu_{\rm c}$ for May 21.0 UT.
           \label{fi:spec}
           }  
\end{figure}

From the light curve behavior and broad-band spectrum
(Fig.~\ref{fi:spec}) of GRB\,970508, Galama et ~al.\
(1998a,b)\nocite{gwbgs:98,gwbgs2:98} deduced the other properties of the
burst required to calculate the physical parameters of the afterglow. We
summarize them here: at $t=12.1\,$d after trigger, the break frequencies
are $\nua=2.5\times10^9\un{Hz}{}$, $\num=8.6\times10^{10}\un{Hz}{}$,
and $\nuc=1.6\times10^{14}\un{Hz}{}$. The peak flux is $F_{\num}=1.7$\,mJy
and the electron index $p=2.2$. After the first 500\,s electrons no longer
cooled efficiently and the afterglow must evolve adiabatically. We
shall set the cosmological parameters to be $\Omega=1$, $\Lambda=0$,
$H_0=70\un{km}{}\un{s}{-1}\un{Mpc}{-1}$. As noted above, they only enter
the solution via the luminosity distance, and alternatives can therefore
be incorporated easily via the substitution given below \Eq{eq:n}.
Finally, we adopt $X=0.7$ for the composition of the ambient medium. There
are no reasons in the model to assume the ambient medium would not have
normal cosmic abundance. While the metallicity $Z$ is a strong function
of redshift, $X$ is hardly redshift-dependent, since the balance between H
and He in cosmic matter has not been changed very much by nucleosynthesis.
Using further that $x_{2.2}=0.580$, $\phi_{2.2}=0.611$, we find
\begin{eqnarray}
    \efiftwo = 3.5  && n = 0.030 \nonumber \\
    \epse = 0.12    && \epsB = 0.089. 
\end{eqnarray}
We do note once more our deliberate use of $\efiftwo$, the energy
per unit solid angle scaled to that of an isotropic explosion of
$10^{52}\un{erg}{}$, in stead of the total energy: $\efiftwo$ is truly
constrained by the data, whereas the total energy requires us to know
the as yet poorly constrained beaming of bursts.  The recent findings
by Fruchter et~al.\ (1999)\nocite{fpgtf:99} suggest there might be
a break in the late light curve (100--200 days after the burst), at
which time the Lorentz factor is 2 or less. If this break is due to
beaming, it would be very modest beaming, and the total energy would
be only a factor of a few less than the isotropic estimate.  Our value
of $\efiftwo$ does clearly rule out the very high energy estimates by
Brainerd (1998)\nocite{brain:98} from the radio data alone. We have
demonstrated for the first time that the electron and magnetic field
energy densities are indeed close to but somewhat below equipartition.
The ambient density is on the low side of normal for a disk of a galaxy
but definitely higher than expected for a halo, lending further support
to the notion that bursts occur in gas-rich environments. As an aside,
we note that switching the values of $\num$ and $\nuc$, which is allowed
by the shape of the spectrum at 12.1\,d, does not give a sensible
solution (e.g., $\epse=20$.) This confirms the choice of Galama et al.\
(1998b)\nocite{gwbgs2:98}, who noted that this solution was not compatible
with the temporal evolution of the afterglow.

The gamma-ray fluence of GRB\,970508 was measured with BATSE to be
$(3.1\pm0.2)\times10^{-6}\un{erg}{}\un{cm}{-2}$. 
Using $z=0.835$ and $h_{70}=1$, this
implies $\efiftwo_\gamma=0.63$. In other words, the energy emitted in
gamma rays is 18\% of the total blast wave energy (per unit solid angle
in our direction). According to Galama
et al.\ (1998b)\nocite{gwbgs2:98}, the afterglow was cooling
efficiently until 500\,s after trigger; this means that during the
gamma-ray phase all the energy given to electrons would be radiated
away quickly, and mostly in gamma rays. If this phase is not too long,
the energy radiated
in gamma rays should be $\efiftwo_\gamma=\epse_\gamma\efiftwo_{\rm
i}$, where $\epse_\gamma$ is the value of $\epse$ during the early,
gamma-ray emitting phase and $\efiftwo_{\rm i}$ is the initial value
of \efiftwo.  Since the subsequent phase will be
adiabatic, the blast wave energy measured at late times should be
$\efiftwo=(1-\epse_\gamma)\efiftwo_{\rm i}$. Eliminating the initial
energy, we conclude that
\be{eq:epsegam}
    \frac{\epse_\gamma}{1-\epse_\gamma}=\frac{\efiftwo_\gamma}{\efiftwo}.
\ee
Therefore the measured ratio of gamma-ray fluence to late time blast wave
energy implies that $\epse_\gamma=0.15$, or slightly greater if some of the
initial energy output is at $E<20\,$keV. Compared with $\epse=0.12$ at
late times, this demonstrates the near-constancy of the fraction of
energy that is given to the electrons. Since the inferences about the
initial gamma-ray fluence are independent of the whole machinery on
blast wave synchrotron emission in the previous section, we may view this
agreement as modest evidence that the coefficients derived there are close
to correct, despite our simplification of the dynamics.

It is also interesting to compare the properties at late times with
those derived from radio observations. The scintillation size after
1 month is about $10^{17}$\,cm (Frail et al.\ 1997)\nocite{fknft:97},
whereas our formulae give a transverse diameter of $5\times10^{17}$\,cm;
given the statistical nature of the scintillation size and our neglect of
the gradients in properties in the transverse direction, to which this
particular measurement is of course sensitive, this is not too bad. The
Lorentz factor at this time is 3.4, so the evolution is still just in the
ultrarelativistic regime. The field at this time is $B^\prime=0.07$\,G.
Katz and Piran (1997)\nocite{kp:97} estimated a size of the afterglow of
GRB\,970508 from a crude measurement of the self-absorption frequency.
They found a size of $10^{17}$\,cm, and assuming an ambient density of
1$\un{cm}{-3}$ they found that the Lorentz factor had already decreased
to 2, and that most of the energy of the blast wave had been lost,
i.e.\ it had evolved with radiative dynamics. The numbers we derive
from our full solution after 1 week are: $\gamma=5.8$, transverse
diameter $=2\times10^{17}$\,cm.  This means the blast wave is still
quite relativistic, and with our low ambient density there is no need
for radiative evolution.

   \section{Properties of GRB\,971214}
   \label{971214}

This burst occurred on 1997 December 14.9727 UT. With a fluence of
$1.1\times10^{-5}\un{erg}{}\un{cm}{-2}$ it is a moderately bright burst
(Kippen et al. 1997)\nocite{kwcsl:97}.  After localization by the BeppoSAX
Wide Field Camera in X rays (Heise et al.\ 1997)\nocite{hzstr:97},
the optical afterglow of this burst was found by Halpern et al.\
(1998)\nocite{hthc:98}. It shows evidence of strong reddening (Halpern et
al. 1998, Ramaprakash et al.\ 1998)\nocite{hthc:98,rkfkk:98}. Once the
afterglow had faded, a host galaxy became visible underneath it and its
redshift was measured to be 3.42 (Kulkarni et al.\ 1998)\nocite{kdrgb:98}.

\begin{figure}[ht]
\centerline{\psfig{figure=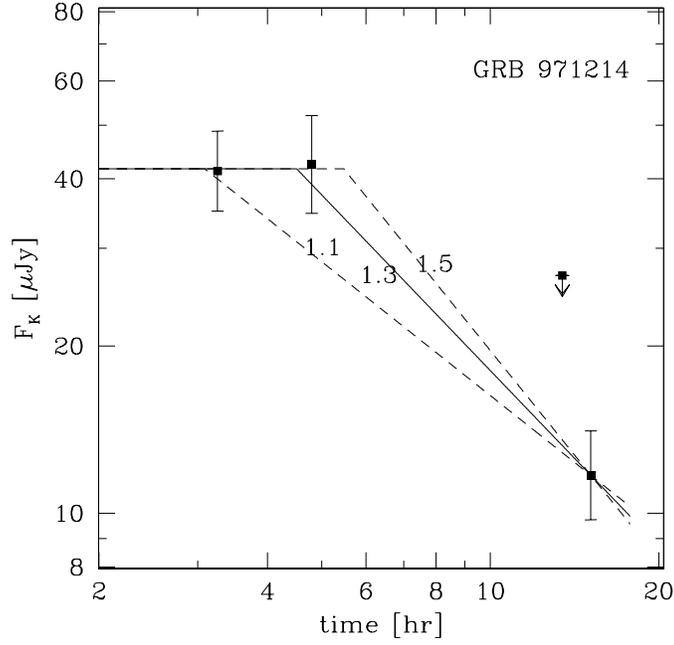,width=8.8cm}}
\caption[]{The $K$ flux of GRB\,971214 as a function of time. Extrapolations
           to early times of the Keck measurement after 14 hours 
	   (Kulkarni et al.\ 1998) are shown for the same three values
	   of $\delta$ used in \Fig{fi:sp971214}. Other data are by 
	   Gorosabel et al.\ (1998) and Garcia et al.\ (1998; the upper
	   limit)\nocite{gmtcm:97}. The assumption here
	   is that $K$ lies above $\num$ and
	   below $\nuc$ (the situation in the righthand panel of
	   \Fig{fi:sp971214}). Therefore, the actual slope
	   of each curve is 0.25 less than the value of $\delta$ (by which
	   it is labelled) because $K$ lies below the cooling break, and
	   $\delta$ is measured in $VRI$, above the cooling break, where
	   the temporal slope is steeper by 1/4 than below it.
	   \label{fi:kband971214}
	   }
\end{figure}

One definite break was observed in the spectrum, and another possible one.
The definite break (henceforth `optical break') was found by Ramaprakash
et al.\ (1998)\nocite{rkfkk:98}; they noted a break in the spectrum of the
afterglow at $3\times10^{14}\un{Hz}{}$, 0.58 days after trigger, in the
extinction-corrected $VRIJK$ spectrum. Another possible break (henceforth
`IR break') was found by Gorosabel et~al.\ (1998)\nocite{gcwhg:98} in $K$
band data 3--5 hours after trigger. In \Fig{fi:kband971214} we show the
$K$ band light curve. The data are not strongly inconsistent with a
pure power law fit ($\chi^2$/dof=2.0), but are suggestive of a break
passing through $K$ after about 5 hours. The physical interpretation of
the afterglow depends rather strongly on whether the optical break
is the peak frequency $\num$ or the cooling frequency $\nuc$, so we shall
discuss these two cases with their implications and problems in turn.

   \subsection{The optical break as $\num$}
   \label{971214.num}

\begin{figure}[ht]
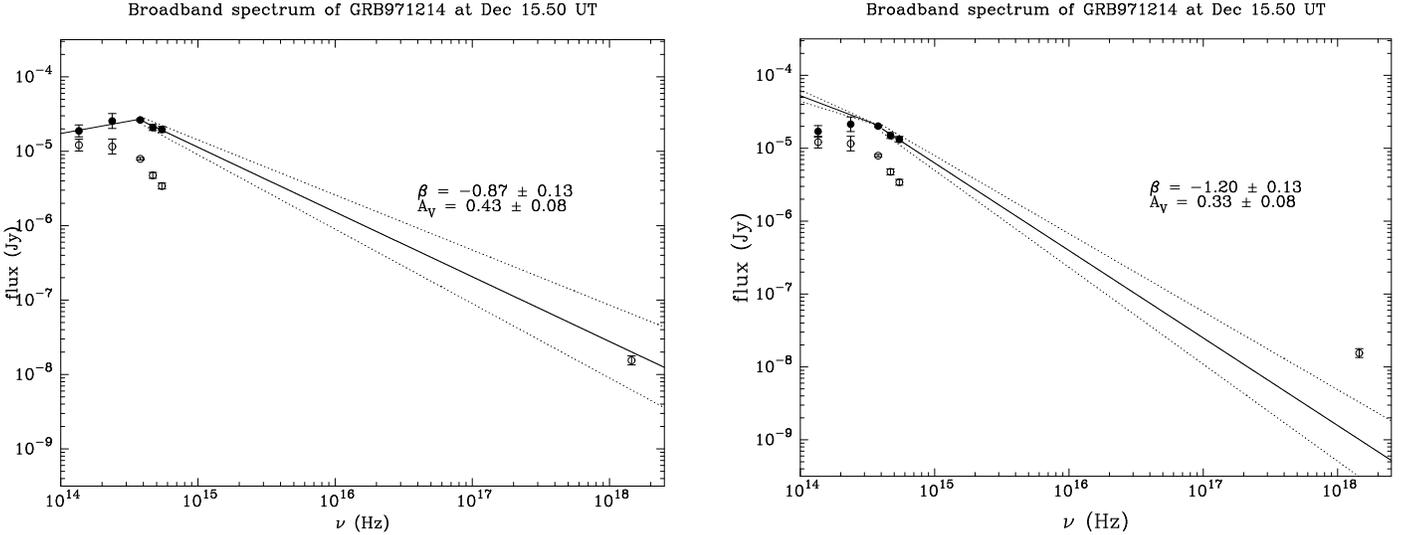

\begin{minipage}[b]{8.8cm}
   \psfig{figure=may8_4l.ps,angle=-90,width=8.8cm}
\end{minipage}\hfill\begin{minipage}[b]{8.8cm}
   \psfig{figure=may8_4r.ps,angle=-90,width=8.8cm}
\end{minipage}
\caption[]{The near-infrared/optical to X-ray spectral flux distribution of
GRB\,971214 on Dec 15.50 UT (the epoch of the $I$ band
measurement). Open symbols indicate the measured values (from
Ramaprakash et al.\ 1998 and Heise et al.\ 1999) extrapolated to Dec
15.50 UT. Note that the error on the J band data is much larger than
used by Ramaprakash et al.\ (1998), in agreement with the original
report (Tanvir et al.\ 1997).  Filled symbols indicate the dereddened
data.  The dotted lines indicate the 1-$\sigma$ errors on the
spectral slope as derived from the temporal slope. 
(left) Result of assuming that the break is the synchrotron peak.
The spectrum below the peak
follows the low-frequency tail of the synchrotron spectrum, where
$F_\nu \propto \nu^{1/3}$. (right) Result of assuming that the break
is the cooling break. The spectral slope changes by 0.5 across the break.
The extinction, $A_V$,
derived from the fit, corresponds to the rest frame $V$ band.
	   \label{fi:sp971214}
	   }
\end{figure}

Ramaprakash et al.\ (1998)\nocite{rkfkk:98} interpreted the optical break
as the peak frequency, $\num$. A complication with the data is that the
spectral slope is much too steep to correspond to any simple fireball
model, which can be interpreted as due to reddening within the host galaxy
(Ramaprakash et al.\ 1998, Halpern et al.\ 1998)\nocite{kdrgb:98,hthc:98}.
Since reddening scales approximately as 1/wavelength, it cannot
be determined without knowing what the true slope of the spectrum
is. Assuming a blast wave model, one can predict this slope from the
temporal decay rate of the flux and an interpretation of what break is
seen. Following Ramaprakash et~al., we now assume an adiabatic blast wave
and interpret the break as $\num$. Then the flux above the break should
depend on frequency and time as $F=F_0\nu^{-\beta}t^{-\delta}$, where
$\beta = 2\delta/3$. There are several determinations of $\delta$, for the
$VRI$ fluxes: $1.2\pm0.2$ by Kulkarni et al.\ (1998)\nocite{kdrgb:98}
and $1.4\pm0.2$ by Halpern et al. (1998)\nocite{hthc:98}. We shall
adopt the value $1.3\pm0.2$ in this paper, so $\beta=0.87\pm0.13$
for the case under consideration. The left hand panel of \Fig{fi:sp971214}
shows the resulting de-reddened spectrum at $t=0.52$\,d (Note that
Ramaprakash et al.\ construct the spectrum at a slightly later time,
0.58\,d). We find that $\num=4\times10^{14}$\,Hz and $F\sub{m}=30\,\mu$Jy.
(As an aside, we note that 
extinction fits with a mean galactic extinction curve are worse, since
the redshifted 2200\,\AA\ bump falls within $VRI$.) The
amount of extinction, 0.43 mag at a rest frame wavelength
of 5500\,\AA, is very modest and does not imply a special location of
the burst within the galaxy. A strong point of this fit is
that the X-ray flux measured at the same time, which was not included in
the fit, agrees nicely with it. The reported non-detections in the
radio at levels of 10--50\,$\mu$Jy could be inconsistent with the peak
flux: the reddening-corrected $F_{\rm m}$ is 30\,$\mu$Jy, so we may have to
invoke self-absorption to suppress the flux at 8.46\,GHz. A weak point
is the fact that the flux at $K$, which is below the peak frequency at
0.58 days in this model, would have to rise with time as $t^{1/2}$ as the
peak approaches. But the early $K$ data by Gorosabel et~al.\ (1998)
clearly indicate a significant decline of the
flux in $K$ from 0.21 to 0.58 days. In fact, the early $K$ flux
even exceeds the supposed peak flux in the spectrum at 0.52\,d, which means
the peak flux had to be declining as well. This requires a non-standard
blast wave model (e.g.\ a beamed one or a non-adiabatic one) and is thus
inconsistent with the blast wave model used in this interpretation.
Nonetheless, we shall briefly explore the physical implications of
this model, since the interpretation of the optical break as a cooling 
break is not free of problems either.

For the simple adiabatic model used by Ramaprakash et al.,
we get the blast wave energy from \Eq{eq:num}:
\be{eq:e971214_num}
   \efiftwo = 27 \frac{2}{1+z} \left(\frac{\epse}{0.2}\right)^{-4}
                 \left(\frac{\epsB}{0.1}\right)^{-1}.
\ee
The coefficient is 60 times larger than in Eq.3 of Ramaprakash et al.,
almost solely due to our more accurate calculation of the
peak frequency.
We can  use \Eq{eq:Fnumobs2}
to derive an independent estimate of the  blast wave energy 
from the peak flux of 30$\mu$Jy:
\be{eq:e971214_Fnum}
   \efiftwo = 0.09\,n^{-1/2} \left(\frac{\epsB}{0.1}\right)^{-1/2}
\ee
This value is difficult to reconcile with the energy estimate from $\num$,
unless we push the equipartition fractions very close to unity and/or
adopt a very low ambient density.

   \subsection{The optical break as $\nuc$}
   \label{971214.nuc}

In order to accomodate the decline in the early $K$ band flux, we now
assume that $\num$ was already well below $K$ band at 0.52\,d, and
therefore the optical break is $\nuc$. Then the spectral
slope in $VRI$ should be related to the temporal decay at those
frequencies by $\beta=(2\delta+1)/3$, which for $\delta=1.3\pm0.2$
implies $\beta=1.2\pm0.13$ and $p=2.4$. This model is explored in the
righthand panel of \Fig{fi:sp971214}. While it solves the $K$ decline
problem, it is a worse fit to the $K$ flux at this time, and does not
do very well in predicting the X-ray flux, which is 2.3$\sigma$ above
the extrapolated spectrum.  Also, if we then say that the IR break is
real, we have a peak flux of 60\,$\mu$Jy. This value is greater than the
8.46\,GHz flux limits obtained 0.8--20 days after the burst (Kulkarni et
al.\ 1998)\nocite{kdrgb:98}.  This would require that the self-absorption
frequency in this afterglow exceeds 10\,GHz. Alternatively,  the peak
frequency at 20 days could be at least a factor 200 above 8.46\,GHz,
so that an extrapolation from the peak flux to the radio using
an optically thin synchrotron spectrum ($F_\nu\propto\nu^{1/3}$)
falls below 10\,$\mu$Jy. Since we know in this case that $\num$ was
$1.4\times10^{14}$\,Hz after 0.21 days, we can use $\num\propto t^{-3/2}$
to find that $\num=150$\,GHz at 20 days, too low to be compatible with the
radio upper limits; we conclude that we must require $\nua\gsim 10$\,GHz,
as in the other interpretation of the break, to suppress the radio flux.

Using $\num$ and $F_{\rm m}$ from
the IR peak at $t=5$\,h$=0.21$\,d, we can again use equations \ref{eq:num}
and \ref{eq:Fnumobs2} to get two expressions 
for the energy in terms of the other unknowns:
\begin{eqnarray}
   \label{eq:e971214_num2}
   \efiftwo & = & 3.0 \left(\frac{\epse}{0.12}\right)^{-4}
                   \left(\frac{\epsB}{0.089}\right)^{-1} \\
   \label{eq:e971214_Fnum2}
   \efiftwo & = & 1.1 \left(\frac{n}{0.030}\right)^{-1/2}
                      \left(\frac{\epsB}{0.089}\right)^{-1/2} 
\end{eqnarray}
Here we have scaled the unknowns to the values found for GRB\,970508. In this
case, the two independent energy estimates are quite compatible.

Now that we identified all but the self-absorption frequency
in the afterglow of GRB\,971214, we may use equations
\ref{eq:efiftwo}--\ref{eq:n} to get all the parameters of the burst,
leaving their dependence on the unknown $\nua$ explicit.
The cooling frequency is $\nuc=4\times10^{14}$\,Hz
at 0.52 days. It follows that
\begin{eqnarray}
    \efiftwo = 0.46\:\nua\sub{,GHz}^{-5/6}  
                 && n = 0.60\:\nua\sub{,GHz}^{25/6} \nonumber \\
    \epse = 0.26\:\nua\sub{,GHz}^{5/6}    
                 && \epsB = 0.027\:\nua\sub{,GHz}^{-5/2}. 
\end{eqnarray}
Since we require $\nua>10$\,GHz in order to satisfy the radio
limits, we get $\efiftwo<0.07$, $n>9\times10^4$, $\epse>1.8$, and
$\epsB<8\times10^{-5}$.  These values are rather different from those of
GRB\,970508, and $\epse>1.8$ is implausible (but not impossible).  With
such a low energy and high density, the GRB would become non-relativistic
(and start a faster decline) within about 4 days. This means that the
late time radio flux may be suppressed without invoking $\nua>10$\,GHz,
easing the constraints on the parameters somewhat. But with $\nua=1$\,GHz
we get values for which the GRB remains relativistic for 400 days; this
conflicts with the radio limits. We conclude that $\nua$ probably has
to exceed 5\,GHz for a consistent solution, implying a fairly low energy
and high ambient density for GRB\,971214. With a high emitted gamma-ray
energy per unit solid angle, $\efiftwo_\gamma=30$, this means that the
ratio of emitted gamma-ray energy to remaining energy in the blast wave
is uncomfortably large, probably in excess of 100. Either very efficient
radiation in the GRB phase of the burst is needed to achieve this, or a strong
difference in the amount of beaming between the gamma-ray and afterglow
emission. Beaming of GRB\,971214 to a degree similar to that observed
in GRB\,990123 (Fruchter et~al.\ 1999, Kulkarni et~al.\ 
1999)\nocite{ftmsp:99,kdobg:99}
would allow a more standard solution for the parameters: all the trouble
arises from the fact that the radio limits force a high $\nua$. If the
burst was beamed so that it started declining more sharply after a day
or so, then the radio flux at later times could be naturally low, and
we would be allowed to choose $\nua\sim1$\,GHz, which leads to parameter
values for GRB\,971214 that are similar to those of 970508. It would even
open the possibility that the temporal decay is somewhat contaminated
by the steeper decay after the beaming break, so we should have used a
slightly shallower temporal and spectral slope in fitting the spectrum of
GRB\,971214 at 0.52\,d with a cooling break. Then both the X-ray and $K$
flux would agree much better with the fit.  However, this now is a little
too much speculation for the amount of data available, and it may well be
wrong to try to bring the parameters of GRB\,971214 into agreement with
those of GRB\,970508. We already know that afterglows are diverse. For
example, GRB\,980703 had a cooling break in X rays after a day rather
than in the optical (Vreeswijk et~al.\ 1999, Bloom et~al.\ 
1998)\nocite{vgoog:99,betal:98}.
We simply need more well-measured bursts in order
to establish the allowed ranges of parameters like $\epse$ and $\epsB$.

In summary, neither the identification of the optical break as $\num$ nor
as $\nuc$ is without problems, so we should take any derived parameters
for this burst with a grain of salt. But the $\num$ interpretation is
in our view the least tenable, since it predicts that the flux at $K$
should have risen from the start until about a day after the burst,
whereas the data clearly show a declining $K$ flux. The problem of
the $\nuc$ interpretation is that the X-ray flux at 0.52 days is about
2.3$\sigma$ higher than expected, and the $K$ band flux lower, which is
somewhat uncomfortable but perhaps tolerable.

   \section{Conclusion}
   \label{conclu}

We have calculated the synchrotron spectra from the blast waves causing
GRB afterglows and derive improved expressions for the relations
between measured break frequencies and the intrinsic properties of
the blast wave. These allow us to relate the blast wave properties to
observable quantities more accurately. We correct the expression for the
blast wave energy by almost two orders of magnitude. Our expressions are
exact for an undecelerated, uniform medium.  Deceleration and radial
structure of the shock are expected to change the expressions for the
final parameters by another factor few, much less than the corrections
found here but still of interest. Combined with the uncertainties in
the measured values of the spectral breaks and fluxes, this means that
the blast wave parameters derived here are still uncertain by an order
of magnitude (see the solution by Granot, Piran, and Sari
1998\nocite{gps:98} as
an illustration of the possible differences).

There is enough data on GRB\,970508 to compute all intrinsic
parameters of the blast wave. The energy in the blast wave is
$3\times10^{52}\un{erg}{}/4\pi\un{sr}{}$. The ambient density into
which the blast wave expands is 0.03$\un{cm}{-3}$, on the low side for
a disk of a galaxy.  The fraction of post-shock energy that goes into
electrons is 12\%, and that into magnetic field, 9\%.  We also estimate
the fraction of energy transferred to electrons during the gamma-ray
phase, and find this to be 15\%. The agreement with the later blast
wave value suggests that the fraction of energy given to electrons is
constant from 10\,s to $10^6$\,s after the trigger.

For GRB\,971214 there is ambiguity in the interpretation of the break
seen in the optical half a day after the burst. We argue that the break is
most likely to be the cooling break, but the argument is not watertight.
Assuming it is the cooling break, we still lack the self-absorption
frequency, but radio limits constrain this to be in excess of about
5\,GHz.  The limits on parameters that follow from this indicate
that the afterglow properties of GRB\,971214 are different from those of
GRB\,970508. GRB\,971214 must either have had more narrow beaming in gamma
rays than in optical or have radiated its initial energy with more than 99\%
efficiency in the gamma-ray phase, according to the parameters we derive.
Also its magnetic field was far below equipartition, $\epsB\lsim10^{-4}$.
However, beaming or other additions could ease the constraints, and allow
parameter values similar to those of GRB\,970508, so
the physical parameters of GRB\,971214 are very poorly constrained.

Our analysis emphasizes the importance of early measurements covering
a wide range of wavelengths. The full identification of the cooling
frequency $\nuc$ in GRB\,970508 hinged on abundant photometry, including
colors, being available soon after the burst, since the break
passed $R$ after 1.5 days (Galama et~al. 1998b\nocite{gwbgs2:98}). In $H$
and $K$, the action lasted a week (Galama et~al. 1998b\nocite{gwbgs2:98}),
and this is the general trend: there is more time in IR, since all
breaks pass later there. However, our revised coefficient for the
peak frequency, $\num$, shows that the peak can only be caught in the
IR within hours of the trigger (or much later in the radio).  A case
in point are the very early $K'$ band measurements of GRB\,971214 by
Gorosabel et al.\ (1998)\nocite{gcwhg:98}, which provide an invaluable
constraint on this afterglow as they may have caught the passage of
$\num$ through $K'$.  Therefore, we encourage first and foremost early
long-wavelength coverage, including searches for afterglows in IR, as
a method of effectively constraining afterglow parameters. Two of the
three crucial break frequencies in an afterglow can pass the optical and IR
within hours and days, respectively.  There is no time to first search
and only then attempt broad coverage.  Instantaneous alerts from HETE2
and SWIFT will therefore greatly advance our understanding of afterglow
physics. For HETE2, and to some extent for SWIFT, this will require an
enormous amount of work from a network of ground-based observatories
with good coverage in longitude and latitude, so that always at least
one observatory is well-placed for immediate response.


\acknowledgments

R.A.M.J. Wijers was supported for part of the work by a Royal Society
URF grant.  T.J. Galama is supported through a grant from NFRA under
contract 781.76.011.  We thank Jules Halpern for alerting us to the
X-ray data on GRB\,971214 and pointing out that our original model was
inconsistent with the X-ray flux.


\newpage

\end{document}